\documentclass[a4paper,12pt]{article}

\begin{document}
\author{Burak Tevfik Kaynak${\,}^{1}$ and  O. Teoman Turgut${\,}^{1,2}$\\  \\ ${}^{1}\,
$Department of Physics, Bogazici University \\
34342 Bebek, Istanbul, Turkey\\ kaynakb@boun.edu.tr\\
and\\ ${ }^{2}$Feza Gursey Institute\\
Kandilli 34684, Istanbul, Turkey\\ turgutte@boun.edu.tr}
\title{\bf Symbol calculus and zeta--function regularized determinants}
\maketitle
\begin{abstract}
In this work, we use semigroup integral to evaluate zeta--function
regularized determinants. This is especially powerful for
non--positive operators such as the Dirac operator. In order to
understand fully the quantum effective action one should know not
only the potential term but also the leading kinetic term. In this
purpose we use the Weyl type of symbol calculus to evaluate the
determinant as a derivative expansion. The technique is applied
both to a spin--$0$ bosonic operator and to the Dirac operator
coupled to a scalar field.
\end{abstract}
\section{Introduction}
Calculation of functional determinants is very important in
quantum field theories. From the one loop effective action to
instanton calculations the main tool is the evaluation of such an
infinite dimensional determinant \cite{ramond,sweinberg}. In this
work, we present a derivative expansion for such regularized
determinants which is especially suitable for non--positive
definite operators, such as the Dirac operator. The literature on
regularized determinants is vast, we will not be able to do
justice to all who has contributed to this area. The main tool is
the introduction of a zeta--function for the operator \cite{seeley,singer1,singer2,hawking}. In quantum
field theory the calculation of the zeta function through the use
of heat kernel, (or its similar version proper time
regularization) is favored \cite{ramond,ball,camporesi,bytsenko,zinn1}, the advantage
is that there is a systematic short time expansion of the heat
kernel, the coefficients of which are all related to geometric
invariants and especially suitable for theories which involve
gauge fields, the disadvantage of this approach is that the
operator under consideration should be positive definite, or its
determinant should be related to the determinant of its square
without any correction terms (i. e. without a multiplicative
anomaly). In general, the regularized determinants do not meet
this last criterion, there is for example by now the well known
Kontsevich--Vishik multiplicative correction
\cite{kontsevich,elizalde,zerbini}. An alternative path is to
evaluate the zeta function through the semigroup integral, which
is used to define complex powers of elliptic operators
\cite{seeley,shubin,saravi1}. In general for higher dimensional
determinants there is no analog of the Gelfand-Yaglom formula
\cite{gelfand} (see however the recent attempts \cite{kirsten1,kirsten2,dunne}). For
such determinants one should resort to an approximation method. It
is physically reasonable to assume that the contributions coming
from the derivatives of the fields are becoming smaller as the
order of the derivative increases. Thus it is natural to look for
a kind of derivative expansion. The proper mathematical tool for
this is the symbol calculus for pseudo--differential operators
\cite{hormander,folland}. In this work we will apply this
expansion to the zeta function via the semigroup integral
representation. In the case of Laplace operators defined over a ball or over a generalized cone, one may actually evaluate the semigroup integral and find an exact result for the determinant \cite{kirsten3,kirsten4}. There are also other cases such as torus $T^N$, sphere $S^N$ and hyperbolic space $H^N$ in which it is possible to find exact solutions of the heat kernel equation and to give the zeta function for the Laplace--Beltrami operator in closed form \cite{bytsenko}. Other exact solutions on homogeneous spaces can be found in \cite{camporesi} as well. It is also possible to give a complete description of the zeta determinants for Dirac and Laplace--type operators over finite cylinders using countour integration method equipped with different boundary conditions \cite{kirsten5}. An important example of the evaluation of chiral
Jacobians via the zeta function method and the symbol calculus is
given in \cite{saravi2}. There are other ways of applying the
symbol calculus essentially exploiting Wigner type transformations
\cite{salcedo} or utilizing a suitable representation of the
logarithm as integral of a resolvent \cite{langmann}, however they
are harder to generalize to manifolds.

In Section \ref{2ndsec} we, first, summarize the well--known
zeta--function prescription for the evaluation of determinants of
an operator, the pseudo--differential operator techniques, and
symbol calculus. Afterwards the definition of complex powers of
elliptic operators via the so called semigroup integral
representation is defined. At the end we conclude that section
with how a regularized determinant of an elliptic operator can be
evaluated by means of introducing a semiclassical expansion for
the symbol of complex powers of the operator under consideration.

In Section \ref{3rdsec} the application of this method throughout
for a spin--$0$ bosonic operator in $4d$--spacetime is explicitly
shown. In Section \ref{4thsec} we elaborate on how a
zeta--function regularized determinant for a Dirac operator with a
scalar field can be calculated. The ways of dealing with some of the techinical difficılties encountered during the calculations are
explained in detail.

In Section \ref{5thsec} the result of Section \ref{4thsec} is
applied to the large--$N$ Yukawa theory and some comments about
the form of the terms involving nonlocal functional of the scalar
field in the $N=\infty$ quantum effective action of the theory are
made.
\section{Zeta function regularization and symbol calculus}\label{2ndsec}
The zeta function of an operator $A$ is defined by the sum over its eigenvalues
\begin{equation}
\zeta(s|A) = \sum_n \frac{1}{\lambda^s_n}.
\end{equation}
This sum converges only for sufficiently large values of $\Re(s)>0$. We introduce a local zeta function by
\begin{equation}
\zeta(s|A)(x) = \left\langle x | A^{-s} | x \right\rangle \,,
\end{equation}
which is a regular analytic function on the complex $s$--plane, otherwise it is possible to define it by means of analytic continuation into the complex $s$--plane. It is known that it is a holomorphic function of $s$ for $\Re(s) > \dim(M)/m$, in which $m$ is the order of the elliptic operator under consideration, and it has a meromorphic extension to the whole complex plane with merely simple poles, and $\zeta(s|A)$ and its derivative are regular at $s=0$ \cite{seeley}. We can calculate the derivative of $\zeta(s|A)$ at $s=0$ as,
\begin{eqnarray}
-\left. \frac{\partial}{\partial s} \zeta(s|A) \right\vert_{s=0} &=& \left. \sum_n \ln \lambda_n e^{-s \ln \lambda_n} \right\vert_{s=0} = \sum_n \ln \lambda_n \\
&=& \ln \prod_n \lambda_n = \ln \det A \, ,
\end{eqnarray}
which allows us to define the regularized determinant. It is assumed that the operator has no eigenvalues near the point zero in order to avoid some infrared divergences. However, it is posible to introduce a restricted determinant by removing these zero modes if the number of zero modes is finite \cite{kirsten1,kirsten2,schwarz}.

Pseudo--differential operators or Calderon--Zygmund operators \cite{seeley,shubin} can be viewed as a generalization of differential operators;
\begin{equation}
A = \sum_{| \alpha | \le m} A_\alpha D^\alpha \quad ( \mathrm{with} \, D^\alpha  = \prod^n_{i=1} (-i\partial / \partial x_i)^{\alpha_i}) \quad \mathrm{and} \quad | \alpha | = \sum_{i=1}^n \alpha_i) \, .
\end{equation}
The symbols are basically smooth matrix--valued functions on the phase space $R^n \oplus R^n$ and can be viewed as a generalization of the characteristic polynomial. In order to set up a one to one correspondence between functions on the phase space and operators acting on the Hilbert space $L^2(R^n)$, one could use Weyl ordering \cite{hormander,folland,rajeev}.

From the action of an operator $A$, its kernel can be easily read,
\begin{equation}
A u(x) = \int dy A(x,y) u(y) \,.
\end{equation}
The symbol, $\tilde{A}(x,p)$, of an operator $A$ can be defined by the Fourier transform of its kernel with respect to the relative coordinate, such that it is given by
\begin{eqnarray}
\tilde{A}(x,p) &=& \int d \xi^d A\left(x + \frac{\xi}{2},x - \frac{\xi}{2}\right) e^{-i \xi \cdot p} \\
A(x,y) &=& \int \frac{d^d p}{(2 \pi)^d} \tilde{A} \left( \frac{x+y}{2},p\right) e^{i p \cdot (x-y) } \, .
\end{eqnarray}

Since the symbols are the functions of the coordinates, $x$ and $p$, the multiplication between the operators will induce a new multiplication rule between their symbols, preserving the multiplication rule. When the multiplication of operators on the Hilbert space is translated into the multiplication of the symbols, we find a noncommutative multiplication which can be given in closed form as
\begin{equation}
\tilde{A} \circ \tilde{B} = \left[ e^{ \frac{i\hbar}{2} \left( \frac{\partial}{\partial x^\mu} \frac{\partial}{\partial p'_\mu} - \frac{\partial}{\partial p_\mu} \frac{\partial}{\partial x'^\mu} \right)} \tilde{A} (x,p) \tilde{B} (x',p')\right]_{x=x' \,; \, p=p'} \, .
\end{equation}
Another way of computing this multiplication is to use so called the generalized Poisson brackets with respect to the phase space coordinates $x$ and $p$ \cite{rajeev}. After expanding the exponential, one ends up with a series consisting of these brackets
\begin{equation}
\tilde{A} \circ \tilde{B} = \sum^\infty_{n=0} \left( \frac{i\hbar}{2}\right)^n \frac{1}{n!} \left\lbrace \tilde{A} , \tilde{B} \right\rbrace_{(n)} \, ,
\end{equation}
where the generalized Poisson brackets are given by
\begin{equation}
\left\lbrace \tilde{A} , \tilde{B} \right\rbrace_{(n)} = \sum^n_{i=0} (-1)^n \tilde{A}^{\nu_1 \cdots \nu_i}_{\mu_1 \cdots \mu_{n-i}} \tilde{B}^{\mu_1 \cdots \mu_{n-i}}_{\nu_1 \cdots \nu_i} \, ,
\end{equation}
in which $\tilde{A}^{\mu_i} = \frac{\partial \tilde{A}}{\partial p_{\mu_i}}$ and $\tilde{A}_{\mu_i} = \frac{\partial \tilde{A}}{\partial x^{\mu_i}}$.

It is, therefore, possible to do a semiclassical expansion with the assistance of this multiplication since after the leading order, which is actually the pointwise multiplication of the operators, the next orders give the desired corrections.

The trace of an operator basically transforms into a phase space integral
\begin{equation}
\mathrm{Tr} A = \int d^d x \frac{d^d p}{(2 \pi)^d} \tilde{A} (x,p) \,.
\end{equation}
If the operator under consideration has discrete indices, then one should also take another trace over these indices, as in the case of Dirac operators which will be discussed in Section \ref{4thsec}.

In general, in order to compute the regularized determinant of an operator, heat kernel method is used. But for this method to work the operator under consideration should be  positive--definite. If the operator is not positive definite then $AA^\dagger$ or $A^\dagger A$ is used. But in this case there can be an extra term coming from the eta invariance or Seeley-De Witt integral coefficients \cite{elizalde,wojciechowski}. There is also another way of evaluating such a determinants independent of its positive definiteness. For the complex powers of operators, there is a very powerful prescription which is called the semigroup integral representation \cite{seeley,shubin}.

Since the semigroup integral representation depends on the resolvent of the operator, our first task should be to find the symbol of the resolvent. In order to do this, it is a good idea to use the product rule so that the resolvent itself and the corrections to it can be calculated. The symbol of the inverse complex power of an operator $A$ is given by
\begin{equation}
\sigma[A^{-s}] = \frac{\sin \pi s}{\pi} \int^\infty_0 d\lambda \, \lambda^{-s}\sigma[\frac{1}{\lambda+A}] \,.
\end{equation}
This definition is formal since it is thought that the symbol of the resolvent fulfills the convergence requirements. As long as a meromorphic extension of the symbol family $\sigma(z,s)$ to whole complex plane with respect to complex parameter $s$ can be found by means of a suitable analytic continuation of this symbol family in $s$, the contour integral makes sense. The next step is to find the symbol of the resolvent as a semiclassical expansion \cite{seeley,rajeev}. If the resolvent symbol is defined as
\begin{equation}
\tilde{R}(\lambda) = \sigma \left[ \frac{1}{\lambda+A} \right]
\end{equation}
then the symbol should satisfy
\begin{equation} \label{res}
\tilde{R}(\lambda) \circ \sigma[\lambda+A] = 1 \,
\end{equation}
where the resolvent symbol can be expanded in a power series in $\hbar$,
\begin{equation}
\quad \tilde{R}(\lambda) = \sum^\infty_{n=0} \hbar^n \tilde{R}_{(n )}(\lambda) \,.
\end{equation}
This expansion is actually a derivative expansion. Equation (\ref{res}) can be rewritten in terms of the generalized Poisson brackets
\begin{eqnarray}\label{resexp}
\tilde{R}(\lambda) \circ \left(\lambda + \tilde{A}\right) &=& \sum^\infty_{n=0} \left( \frac{i\hbar}{2}\right)^n \frac{1}{n!} \left\lbrace \tilde{R}(\lambda) , \lambda + \tilde{A} \right\rbrace_{(n)} \nonumber \\
&=& \tilde{R}(\lambda) \left(\lambda + \tilde{A}\right) + \left( \frac{i\hbar}{2}\right) \left\lbrace \tilde{R}(\lambda) , \lambda + \tilde{A} \right\rbrace_{(1)} \nonumber \\ & & + \left( \frac{i\hbar}{2}\right)^2 \frac{1}{2!} \left\lbrace \tilde{R}(\lambda) , \lambda + \tilde{A} \right\rbrace_{(2)} \nonumber \\ & &+ \left( \frac{i\hbar}{2}\right)^3 \frac{1}{3!} \left\lbrace \tilde{R}(\lambda) , \lambda + \tilde{A} \right\rbrace_{(3)} + \cdots \, ,
\end{eqnarray}
where $\tilde{A}$ is just the symbol of the operator $A$.
The equation (\ref{res}) results in a set of recursion relations after collecting the terms with the same order in $\hbar$ and we get
\begin{eqnarray}
\hbar^0 &:& \tilde{R}_{(0)} = \frac{1}{\left( \lambda + \tilde{A} \right)} \label{h0} \\
\hbar^1 &:& \tilde{R}_{(1)} = -\frac{i}{2} \left\lbrace \tilde{R}_{(0)}(\lambda) , \lambda + \tilde{A} \right\rbrace_{(1)} \frac{1}{\left( \lambda + \tilde{A} \right)} \label{h1} \\
\hbar^2 &:& \tilde{R}_{(2)} = - \frac{i}{2} \left\lbrace \tilde{R}_{(1)}(\lambda) , \lambda + \tilde{A} \right\rbrace_{(1)} \frac{1}{\left( \lambda + \tilde{A} \right)} + \frac{1}{8} \left\lbrace \tilde{R}_{(0)}(\lambda) , \lambda + \tilde{A} \right\rbrace_{(2)} \frac{1}{\left( \lambda + \tilde{A} \right)} \nonumber \\ \\
\vdots &:& \qquad \qquad \qquad \qquad \qquad \qquad \qquad \qquad \vdots \qquad \qquad \qquad \qquad \qquad \qquad \qquad \qquad \nonumber
\end{eqnarray}
By means of this expansion, one becomes ready to evaluate the symbol of any inverse complex power of the desired operator order by order in $\hbar$ as follows
\begin{eqnarray}
\sigma[A^{-s}] &=& \frac{\sin \pi s}{\pi} \int^\infty_0 d\lambda \, \lambda^{-s} \frac{1}{\lambda+ \tilde{A}} \nonumber \\
& & -\frac{i \hbar}{2} \frac{\sin \pi s}{\pi} \int^\infty_0 d\lambda \, \lambda^{-s} \left\lbrace \tilde{R}_{(0)}(\lambda) , \lambda + \tilde{A} \right\rbrace_{(1)} \frac{1}{\left( \lambda + \tilde{A} \right)} \nonumber \\
& & - \frac{i \hbar^2}{2} \frac{\sin \pi s}{\pi} \int^\infty_0 d\lambda \, \lambda^{-s} \left\lbrace \tilde{R}_{(1)}(\lambda) , \lambda + \tilde{A} \right\rbrace_{(1)} \frac{1}{\left( \lambda + \tilde{A} \right)} \nonumber \\
& & + \frac{\hbar^2}{8} \frac{\sin \pi s}{\pi} \int^\infty_0 d\lambda \, \lambda^{-s} \left\lbrace \tilde{R}_{(0)}(\lambda) , \lambda + \tilde{A} \right\rbrace_{(2)} \frac{1}{\left( \lambda + \tilde{A} \right)} \nonumber \\
& & + \cdots \,.
\end{eqnarray}
The next step is to take the phase space integral in order to find the zeta function of the operator,
\begin{equation}
\zeta ( s|A ) = \int d^d x \frac{d^d p}{(2 \pi)^d} \sigma[A^{-s}] \,.
\end{equation}
As mentioned at the beginning of this section, the regularized determinant of the operator is just minus the derivative of the zeta function with respect to the complex parameter $s$ at $s=0$, therefore one finds,
\begin{equation}
\ln \det (A) = - \left. \frac{\partial}{\partial s} \right\vert_{s=0}\ \int d^d x \frac{d^d p}{(2 \pi)^d} \sigma[A^{-s}] \,.
\end{equation}
\section{The determinant of a bosonic operator}\label{3rdsec}
For the bosonic case with zero spin, the operator which we would like to evaluate its zeta function is $A = -\partial^2+V(x)$ where the term $V(x)$ may actually be a functional of the field $\phi(x)$. For example in the massless $\phi^4$--theory, the potential term is merely $\lambda \phi^2(x)/2$. The first Poisson bracket is
\begin{eqnarray}
\left\lbrace \tilde{R}_{(0)}(\lambda) , \lambda + \tilde{A} \right\rbrace_{(1)} &=& \frac{\partial}{\partial x^\mu} \frac{1}{\left( \lambda + \tilde{A} \right)} \frac{\partial}{\partial p_\mu} \left( \lambda + \tilde{A} \right) -
\frac{\partial}{\partial p_\mu} \frac{1}{\left( \lambda + \tilde{A} \right)} \frac{\partial}{\partial x^\mu} \left( \lambda + \tilde{A} \right) \nonumber \\
 &=& 0
\end{eqnarray}
The first correction to the resolvent is, thus, zero
\begin{equation}
\tilde{R}_{(1)}(\lambda) = 0 \,.
\end{equation}
The second generalized Poisson bracket is given by
\begin{eqnarray}
\left\lbrace \tilde{R}_{(0)}(\lambda) , \lambda + \tilde{A} \right\rbrace_{(2)} &=& \frac{\partial^2}{\partial x^\mu \partial x^\nu} \frac{1}{\left( \lambda + \tilde{A} \right)} \frac{\partial^2}{\partial p_\mu\partial p_\nu} \left( \lambda + \tilde{A} \right) \nonumber \\
& & -\frac{\partial^2}{\partial x^\mu \partial p_\nu} \frac{1}{\left( \lambda + \tilde{A} \right)} \frac{\partial^2}{\partial x^\nu \partial p_\mu} \left( \lambda + \tilde{A} \right) \nonumber \\
& & + \frac{\partial^2}{\partial p_\mu\partial p_\nu} \frac{1}{\left( \lambda + \tilde{A} \right)} \frac{\partial^2}{\partial x^\mu \partial x^\nu} \left( \lambda + \tilde{A} \right) \nonumber \\
 &=& \frac{4}{\left( \lambda + \tilde{A} \right)^3} \frac{\partial \tilde{V}}{\partial x^\mu} \frac{\partial \tilde{V}}{\partial x_\mu} + \frac{8 p^\mu p^\nu}{\left( \lambda + \tilde{A} \right)^3} \frac{\partial^2 \tilde{V}\tilde{V}}{\partial x^\mu \partial x^\nu} - \frac{4}{\left( \lambda + \tilde{A} \right)^2} \frac{\partial^2 \tilde{V}}{\partial x^\mu \partial x_\mu}
\,. \nonumber \\
\end{eqnarray}
The first nonzero contribution is
\begin{equation}
\tilde{R}_{(2)}(\lambda) = - \frac{4}{\left( \lambda + \tilde{A} \right)^3} \frac{\partial^2 \tilde{V}}{\partial x^\mu \partial x_\mu} + \frac{4}{ \left( \lambda + \tilde{A} \right)^4} \frac{\partial \tilde{V}}{\partial x^\mu} \frac{\partial \tilde{V}}{\partial x_\mu} + \frac{8 p^\mu p^\nu}{\left( \lambda + \tilde{A} \right)^4} \frac{\partial^2 \tilde{V}}{\partial x^\mu \partial x^\nu} \,.
\end{equation}
Therefore, the semiclassical expansion for the resolvent symbol can be given by
\begin{equation}
\tilde{R}(\lambda) = \frac{1}{\left( \lambda + \tilde{A} \right)} - \frac{1}{2 \left( \lambda + \tilde{A} \right)^3} \frac{\partial^2 \tilde{V}}{\partial x^\mu \partial x_\mu} + \frac{1}{2 \left( \lambda + \tilde{A} \right)^4} \frac{\partial \tilde{V}}{\partial x^\mu} \frac{\partial \tilde{V}}{\partial x_\mu} + \frac{p^\mu p^\nu}{\left( \lambda + \tilde{A} \right)^4} \frac{\partial^2 \tilde{V}}{\partial x^\mu \partial x^\nu} + \cdots \,,
\end{equation}
in which $\hbar$ is set to $1$.
The next step is to evaluate the semigroup integrals of this resolvent symbol in order to find the symbol of any complex power of the operator as an expansion,
\begin{eqnarray}
\sigma[A^{-s}] &=& \frac{\sin \pi s}{\pi} \int^\infty_0 d\lambda \, \lambda^{-s} \left[ \frac{1}{\left( \lambda + \tilde{A} \right)} - \frac{1}{2 \left( \lambda + \tilde{A} \right)^3} \frac{\partial^2 \tilde{V}}{\partial x^\mu \partial x_\mu} \right. \nonumber \\ & & + \left. \frac{1}{2 \left( \lambda + \tilde{A} \right)^4} \frac{\partial \tilde{V}}{\partial x^\mu} \frac{\partial \tilde{V}}{\partial x_\mu} + \frac{p^\mu p^\nu}{\left( \lambda + \tilde{A} \right)^4} \frac{\partial^2 \tilde{V}}{\partial x^\mu \partial x^\nu} \right] + \cdots \,.
\end{eqnarray}
All the desired integrals are standard residue integrals and after evaluating these integrals, one ends up with
\begin{equation}
\sigma[A^{-s}] = \frac{1}{\tilde{A}^s} - \frac{s(s+1)}{4\tilde{A}^{s+2}} \frac{\partial^2 \tilde{V}}{\partial x^\mu \partial x_\mu} + \frac{s(s+1)(s+2)}{6\tilde{A}^{s+3}} \left[ p^\mu p^\nu  \frac{\partial^2 \tilde{V}}{\partial x^\mu \partial x^\nu} + \frac{1}{2} \frac{\partial \tilde{V}}{\partial x^\mu} \frac{\partial \tilde{V}}{\partial x_\mu} \right] + \cdots \,.
\end{equation}
Since we found the symbol of the inverse complex power of the operator, taking the trace of the symbol in phase space is left. We take the momentum integral first,
\begin{eqnarray}
\int \frac{d^4 p}{(2 \pi)^4} \sigma \left[ A^{-s} \right] &=& \int \frac{d^4 p}{(2 \pi)^4} \frac{1}{\left(p^2 + \tilde{V} \right)^s} - \frac{s(s+1)}{4} \frac{\partial^2 \tilde{V}}{\partial x^\mu \partial x_\mu} \int \frac{d^4 p}{(2 \pi)^4} \frac{1}{\left(p^2 + \tilde{V} \right)^{s+2}} \nonumber \\
& & + \frac{s(s+1)(s+2)}{6} \frac{\partial^2 \tilde{V}}{\partial x^\mu \partial x^\nu} \int \frac{d^4 p}{(2 \pi)^4} \frac{p^\mu p^\nu}{\left(p^2 + \tilde{V} \right)^{s+3}} \nonumber \\
& & + \frac{s(s+1)(s+2)}{6} \frac{1}{2} \frac{\partial \tilde{V}}{\partial x^\mu} \frac{\partial \tilde{V}}{\partial x_\mu} \int \frac{d^4 p}{(2 \pi)^4} \frac{1}{\left(p^2 + \tilde{V} \right)^{s+3}} + \cdots \nonumber \\
&=& \frac{1}{16 \pi^2} \left[ \frac{\tilde{V}^{2-s}}{(s-1)(s-2)} - \frac{\tilde{V}^{-s}}{6} \frac{\partial^2 \tilde{V}}{\partial x^\mu \partial x_\mu} + \frac{s \tilde{V}^{-s-1}}{12} \frac{\partial \tilde{V}}{\partial x^\mu} \frac{\partial \tilde{V}}{\partial x_\mu} \right] + \cdots \,. \nonumber \\
\end{eqnarray}
Thus the zeta function of the operator $A$ in ordinary spacetime is just the $x$--integral of the equation above,
\begin{equation}
\zeta(s \vert A) = \frac{1}{16 \pi^2} \int d^4 x \, \left[ \frac{V^{2-s}}{(s-1)(s-2)} - \frac{V^{-s}}{6} \frac{\partial^2 V}{\partial x^\mu \partial x_\mu} + \frac{s V^{-s-1}}{12} \frac{\partial V}{\partial x^\mu} \frac{\partial V}{\partial x_\mu} \right] + \cdots \,.
\end{equation}
The determinant of the operator is given by
\begin{equation}
\ln \det \left[ - \partial^2 + V \right] = - \zeta' \left( 0 | - \partial^2 + V \right) \, ,
\end{equation}
and the derivative of the zeta function with respect to $s$ is
\begin{eqnarray}
\frac{\partial}{\partial s} \zeta(s)
&=& \frac{1}{16 \pi^2} \int d^4 x \left[ - \frac{V^{2-s}}{(s-1)^2(s-2)} - \frac{V^{2-s}}{(s-1)(s-2)^2} - \frac{V^{2-s} \ln V }{(s-1)(s-2)} \right. \nonumber \\
& & + \left. \frac{V^{-s} \ln V}{6} \frac{\partial^2 V}{\partial x^\mu \partial x_\mu} +\frac{ V^{-s-1}}{12} \frac{\partial V}{\partial x^\mu} \frac{\partial V}{\partial x_\mu} - \frac{s V^{-s-1} \ln V}{12} \frac{\partial V}{\partial x^\mu} \frac{\partial V}{\partial x_\mu} \right] + \cdots \,. \nonumber \\
\end{eqnarray}
The zeta--regularized determinant of the operator $-\partial^2+V$ is, thus,
\begin{equation}\label{bosdet}
\ln \det \left[ - \partial^2 + V \right] = \frac{1}{32 \pi^2} \int d^4 x \left[  V^2 \ln \left( e^{-3/2} \frac{V}{\mu^{[V]}} \right) + \frac{ 1}{6 V} \frac{\partial V}{\partial x^\mu} \frac{\partial V}{\partial x_\mu} \right] + \cdots \, ,
\end{equation}
where the scale $\mu$ is introduced for dimensional bookkeeping of the logarithm and $[V]$ is the mass dimension of the potential. This result agrees with the ones in the literature \cite{ball,itzykson,chan}. Replacing $V(x)$ with $\lambda \phi^2(x)/2$, which is the potential for the massless $\phi^4$--theory as it is said at the beginning of this section, the potential part of equation (\ref{bosdet}) yields
\begin{equation}
\frac{1}{32 \pi^2} \int d^4 x \frac{\lambda^2 \phi^4}{4} \ln \left( e^{-3/2} \frac{\lambda \phi^2}{2\mu^2} \right) \,.
\end{equation}
This is in agreement with the well known unrenormalized results which can be found in \cite{ramond,eweinberg}.
\section{The determinant of the Dirac operator}\label{4thsec}
In this section we will consider a Dirac operator which is massless and contains a scalar field, $D = \gamma \cdot \partial + \phi$, where $\cdot$ stands for the $4d$--Euclidean inner product and the gamma matrices are chosen to be hermitian. As we have done for the scalar determinant, our first task is to evaluate the resolvent symbol. For the Dirac operator under consideration, the symbol of the operator is
\begin{equation}\label{sd}
\tilde{A} = i \gamma \cdot p + \tilde{\phi} \, .
\end{equation}
After using the same expansion in equation (\ref{resexp}) with the symbol (\ref{sd}), equation (\ref{h0}) and  equation (\ref{h1}) give, respectively, the zeroth and the first order resolvent symbols as
\begin{eqnarray}
\tilde{R}_{(0)}(\lambda) &=& \frac{1}{\left( \lambda + \tilde{A} \right)} \\
\tilde{R}_{(1)}(\lambda) &=& -\frac{i}{2} \frac{\partial \tilde{\phi}}{\partial x^\mu} \left[ - \frac{1}{\left( \lambda + \tilde{A} \right)^2}  i \gamma^\mu + \frac{1}{\left( \lambda + \tilde{A} \right)} i \gamma^\mu \frac{1}{\left( \lambda + \tilde{A} \right)} \right] \frac{1}{\left( \lambda + \tilde{A} \right)} \, .\nonumber \\
\end{eqnarray}
However, when the Dirac operator is considered, there is another trace which is over the spinor indices and if one takes it into account, it can be easily seen that the term in $\hbar$ is zero due to the cyclicity of the trace,
\begin{equation}
\mathrm{Tr} \tilde{R}_{(1)}(\lambda) =0 \, ,
\end{equation}
As in the scalar case, there is not any term which contains just one derivative of the field for the Dirac operator. The next term, which is an $\hbar^2$--order term, contains second derivative of the scalar field and is just given by the second generalized Poisson bracket due to the fact that $\tilde{R}_{(1)}(\lambda)$ is zero,
\begin{eqnarray}
\tilde{R}_{(2)}(\lambda) &=& \frac{1}{8} \left\lbrace \tilde{R}_{(0)}(\lambda) , \lambda + \tilde{A} \right\rbrace_{(2)} \frac{1}{\left( \lambda + \tilde{A} \right)} \nonumber \\
&=& \frac{1}{8} \left[\frac{\partial^2}{\partial p_\mu\partial p_\nu} \frac{1}{\left( \lambda + \tilde{A} \right)} \frac{\partial^2}{\partial x^\mu \partial x^\nu} \left( \lambda + \tilde{A} \right) \right] \frac{1}{\left( \lambda + \tilde{A} \right)} \, .
\end{eqnarray}
This is the only nonvanishing term in the Poisson bracket and it is equal to
\begin{eqnarray}
 && \frac{1}{8} \partial_{\mu \nu} \tilde{\phi} \left[ \frac{1}{\left( \lambda + \tilde{A} \right)} i \gamma^\mu \frac{1}{\left( \lambda + \tilde{A} \right)} i \gamma^\nu \frac{1}{\left( \lambda + \tilde{A} \right)} \right. \nonumber \\
 && + \left. \frac{1}{\left( \lambda + \tilde{A} \right)} i \gamma^\nu \frac{1}{\left( \lambda + \tilde{A} \right)} i \gamma^\mu \frac{1}{\left( \lambda + \tilde{A} \right)}\right] \frac{1}{\left( \lambda + \tilde{A} \right)} \,.
\end{eqnarray}
For the whole expression is multiplied by the second derivative of the field $\tilde{\phi}$, symmetric in $\mu$ and $\nu$, this term simplifies more and after using the cyclicity of the trace, the first nonzero contribution coming from the semiclassical expansion is given by
\begin{equation}
\tilde{R}_{(2)}(\lambda) = - \frac{1}{4}  \frac{1}{\left( \lambda + \tilde{A} \right)^3} \gamma^\mu \frac{1}{\left( \lambda + \tilde{A} \right)} \gamma^\nu \partial_{\mu \nu} \tilde{\phi} \,.
\end{equation}
Therefore, the resolvent symbol up to $\hbar^2$--order is given by
\begin{equation}
\tilde{R}(\lambda) = \frac{1}{\left( \lambda + \tilde{A} \right)} - \frac{1}{4} \partial_{\mu \nu} \tilde{\phi} \frac{1}{\left( \lambda + \tilde{A} \right)^3} \gamma^\mu \frac{1}{\left( \lambda + \tilde{A} \right)} \gamma^\nu +\cdots \, .
\end{equation}

What should be done afterwards is to take the residue integrals in order to find the symbol of the inverse complex power of our operator,
\begin{eqnarray}\label{symdir}
\sigma[A^{-s}] &=& \frac{\sin \pi s}{\pi} \int^\infty_0 d\lambda \, \lambda^{-s} \left[\frac{1}{\left( \lambda + \tilde{A} \right)} - \frac{1}{4} \partial_{\mu \nu} \tilde{\phi} \frac{1}{\left( \lambda + \tilde{A} \right)^3} \gamma^\mu \frac{1}{\left( \lambda + \tilde{A} \right)} \gamma^\nu \right] + \cdots \nonumber \\
&=& \frac{1}{\tilde{A}^s} -\frac{\sin \pi s}{\pi} \int^\infty_0 d\lambda \, \lambda^{-s} \frac{1}{4} \partial_{\mu \nu} \tilde{\phi} \frac{1}{\left( \lambda + \tilde{A} \right)^3} \gamma^\mu \frac{1}{\left( \lambda + \tilde{A} \right)} \gamma^\nu + \cdots \,.
\end{eqnarray}
So as to succeed in taking the second integral, we should reorganize the order of the gamma matrices and the fractions such that the gamma matrices and the fractions should be separated. This can be achieved by means of passing one of the gamma matrices over the fraction between them. By expading the fraction as,
\begin{equation}
\frac{1}{\lambda + \tilde{\phi} + i \gamma \cdot p} = \frac{1}{\lambda + \tilde{\phi}} \sum^\infty_{n=0} (-1)^n \frac{\Gamma(1+n)}{n! \Gamma(1)} \left(\frac{i \gamma \cdot p}{\lambda + \tilde{\phi}} \right)^n \,,
\end{equation}
and using an analytical continuation argument, one can easily show that
\begin{equation}\label{grg}
\gamma^\mu \frac{1}{\lambda+\tilde{A}} \gamma^\nu = \frac{1}{\lambda+\tilde{A}^\ast} \gamma^\mu \gamma^\nu + \frac{1}{\lambda+\tilde{A}} \frac{p^\mu}{p^2} p \cdot \gamma \gamma^\nu - \frac{1}{\lambda+\tilde{A}^\ast} \frac{p^\mu}{p^2} p \cdot \gamma \gamma^\nu \,,
\end{equation}
in which $\tilde{A}^\ast = \tilde{\phi} - i \gamma \cdot p$. After plugging equation (\ref{grg}) into equation (\ref{symdir}), we get
\begin{eqnarray}\label{spd}
 \sigma[A^{-s}] &=& \frac{1}{\tilde{A}^s} -\frac{\sin \pi s}{\pi} \int^\infty_0 d\lambda \, \lambda^{-s} \frac{1}{4} \frac{1}{\left( \lambda + \tilde{A} \right)^3} \frac{1}{\left(\lambda+\tilde{A}^\ast\right)} \gamma^\mu \gamma^\nu \partial_{\mu \nu} \tilde{\phi} \nonumber \\
&& -\frac{\sin \pi s}{\pi} \int^\infty_0 d\lambda \, \lambda^{-s} \frac{1}{4}  \frac{1}{\left( \lambda + \tilde{A} \right)^4} \frac{p^\mu}{p^2} p \cdot \gamma \gamma^\nu \partial_{\mu \nu} \tilde{\phi} \nonumber \\
&& +\frac{\sin \pi s}{\pi} \int^\infty_0 d\lambda \, \lambda^{-s} \frac{1}{4}  \frac{1}{\left( \lambda + \tilde{A} \right)^3} \frac{1}{\left(\lambda+\tilde{A}^\ast\right)} \frac{p^\mu}{p^2} p \cdot \gamma \gamma^\nu \partial_{\mu \nu} \tilde{\phi} \,.
\end{eqnarray}
We start with the first term: the trace, both over continuous indices, $x$ and $p$, and over spinor indices, is going to be taken. In order to do this, it is a good idea to take the momentum integral in $d$--spacetime first and then take the limit as $d\rightarrow4$.
\begin{eqnarray}
\mathrm{tr} \int \frac{d^d p}{(2 \pi)^d} \frac{1}{\left( \tilde{\phi} + i p \cdot \gamma \right)^s} &=& \mathrm{tr} \int \frac{d^d p}{(2 \pi)^d} \phi^{-s} \sum^\infty_{n=0} (-1)^n \frac{\Gamma(s+n)}{n! \Gamma(s)} \left( \frac{i p \cdot \gamma}{\tilde{\phi}} \right)^n \nonumber \\
&=& \frac{2 \tilde{\phi}^{-s}}{(2 \pi)^{d/2} \Gamma(d/2)} \int^\infty_0 dp p^{d-1} {}_2 F_1 \left( \frac{s}{2} , \frac{s+1}{2} ; \frac{1}{2} ; - \frac{p^2}{\tilde{\phi}^2}\right) \nonumber \\
&=& \frac{2^{d/2} \pi^{1/2-d/2} \tilde{\phi}^{-s} \left(\tilde{\phi}^2\right)^{d/2} \Gamma(s-d)}{\Gamma(1/2-d/2) \Gamma(s)} \, , \quad \left| \arg(1/ \tilde{\phi}^2) \right|< \pi \, , \nonumber \\ & & 0< \Re (d/2) < \Re (s/2) < \Re (s/2+1/2) \, ,
\end{eqnarray}
where the little trace stands for the one over the spinor indices and the fact that only even number of gamma matrices is nonvanishing is used. The result of integral in the third line is given in \cite{prudnikov}. As $d\rightarrow4$, the expression above becomes
\begin{equation}
\mathrm{tr} \int \frac{d p^4}{(2 \pi)^4} \frac{1}{\left( \tilde{\phi} + i p \cdot \gamma \right)^s} = \frac{3 \tilde{\phi}^{4-s} \Gamma(s-4)}{\pi^2 \Gamma(s)} \,.
\end{equation}
If one evaluates minus the derivative of this momentum integral with respect to $s$ at $s=0$, then the zeroth order term will be obtained and this is equal to
\begin{equation}
\frac{1}{16 \pi^2} \int d^4 x \phi^4 \ln \left( \frac{\phi^2}{\mu^2} e^{-25/6}\right) \,,
\end{equation}
where the scale $\mu$ is again introduced for dimensional bookkeeping of the logarithm. This agrees with the result given in \cite{salcedo}.

The third term in equation (\ref{spd}) can also easily be calculated in the same manner and one can get
\begin{equation}
-\frac{1}{4} \mathrm{tr} \int \frac{d^4 p}{(2 \pi )^4} \frac{\sin \pi s}{\pi} \int^\infty_0 d\lambda \, \lambda^{-s} \frac{1}{\left( \lambda + \tilde{A} \right)^4} \frac{p^\mu}{p^2} p \cdot \gamma \gamma^\nu \partial_{\mu \nu} \tilde{\phi} = -\frac{1}{32 \pi^2} \frac{\tilde{\phi}^{1-s}}{s-1} \partial^2 \tilde{\phi} \,.
\end{equation}

However, one should be cautious for the second and the fourth terms due to the fact that these terms contain the product of two operators noncommuting with each other. Therefore a method must be suggested to take the residue integrals properly.

The first one is to use the Feynman parametrization so as to convert this multiplication of inverse powers into a single inverse power,
\begin{eqnarray}
 \frac{1}{\left( \lambda + \tilde{A} \right)^3} \frac{1}{\left(\lambda+\tilde{A}^\ast\right)} &=& \mathcal{FP} \int^1_0 dt \frac{3 t^2}{\left[ t ( \lambda + \tilde{A} ) + (1-t) ( \lambda + \tilde{A}^\ast ) \right]^4} \nonumber \\
&=& \mathcal{FP} \int^1_0 dt \frac{3 t^2}{\left[ \lambda + \tilde{\phi} + ( 2 t -1) i p \cdot \gamma \right]^4} \,,
\end{eqnarray}
in which the above integral is given as a Hadamard finite part integral since an extraneous singularity is introduced while this parametrization is being utilized. Thus the second term in equation (\ref{spd}) is
\begin{eqnarray}
-\frac{1}{4} \mathrm{tr} \int \frac{d^4 p}{(2 \pi )^4} \frac{\sin \pi s}{\pi} \int^\infty_0 d\lambda \, \lambda^{-s} \frac{1}{(\lambda +\tilde{A})^3(\lambda +\tilde{A}^\ast)} \gamma^\mu \gamma^\nu \partial_{\mu \nu} \tilde{\phi} \nonumber \\
= - \frac{3}{4} \partial^2 \tilde{\phi} \mathcal{FP} \int^1_0 dt t^2 \, \mathrm{tr} \int \frac{d^4 p}{(2 \pi )^4} \frac{\sin \pi s}{\pi} \int^\infty_0 d\lambda \, \lambda^{-s} \frac{1}{(\lambda + \tilde{\phi} + ( 2 t -1) i p \cdot \gamma)^4} \,,
\end{eqnarray}
where there is just one inverse power in the residue integral. After the residue integral, the expression above becomes
\begin{equation}
- \frac{1}{8} \partial^2 \tilde{\phi} \mathcal{FP} \int^1_0 dt t^2  \frac{\Gamma(s+3)}{\Gamma(s)} \mathrm{tr} \int \frac{d^4 p}{(2 \pi)^4} \frac{1}{ \left( \tilde{\phi} + (2t-1) i p \cdot \gamma \right)^{s+3}} \,,
\end{equation}
which is equal to
\begin{equation}
- \frac{3}{8 \pi^2} \partial^2 \tilde{\phi} \frac{\tilde{\phi}}{s-1} \mathcal{FP} \int^1_0 dt \frac{t^2}{(2t-1)^4}\,.
\end{equation}
Since the Hadamard finite part of the integral is $-1/3$, the second term in the expansion is
\begin{equation}
-\frac{1}{4} \mathrm{tr} \int \frac{d^4 p}{(2 \pi )^4} \frac{\sin \pi s}{\pi} \int^\infty_0 d\lambda \, \lambda^{-s} \frac{1}{(\lambda +\tilde{A})^3(\lambda +\tilde{A}^\ast)} \gamma^\mu \gamma^\nu \partial_{\mu \nu} \tilde{\phi} = \frac{1}{8 \pi^2} \frac{\tilde{\phi}^{1-s}}{s-1} \partial^2 \tilde{\phi} \,.
\end{equation}
If the same calculations are done step by step for the fourth term in equation (\ref{spd}), one gets
\begin{equation}
\frac{1}{4} \mathrm{tr} \int \frac{d^4 p}{(2 \pi )^4} \frac{\sin \pi s}{\pi} \int^\infty_0 d\lambda \, \lambda^{-s} \frac{1}{\left( \lambda + \tilde{A} \right)^3} \frac{1}{\left( \lambda + \tilde{A}^\ast \right)}\frac{p^\mu}{p^2} p \cdot \gamma \gamma^\nu \partial_{\mu \nu} \tilde{\phi} = -\frac{1}{32 \pi^2} \frac{\tilde{\phi}^{1-s}}{s-1} \partial^2 \tilde{\phi} \,.
\end{equation}
The zeta function of the operator up to $\hbar^2$--order is, then,
\begin{equation}
\zeta\left( s | \gamma \cdot \partial + \phi \right) = \frac{1}{\pi^2} \int d^4x \left[  \frac{3 \tilde{\phi}^{4-s}}{(s-1)(s-2)(s-3)(s-4)} + \frac{1}{16} \frac{\tilde{\phi}^{1-s}}{s-1} \partial^2 \tilde{\phi} \right] + \cdots \,.
\end{equation}

The determinant of the Dirac operator with a scalar field $\phi$ up to $\hbar^2$-- order can, therefore, be given by
\begin{equation}
\ln \det (\gamma \cdot \partial + \phi) = \frac{1}{16 \pi^2} \int d^4x \left[ \phi^4 \ln \left( \frac{\phi^2}{\mu^2} e^{-25/6}\right) + \ln \left( \frac{\phi^2}{\mu^2} \right) \frac{1}{2} \partial_\mu \phi \partial^\mu \phi \right] + \cdots \,.
\end{equation}

The alternative way of attacking the second and the fourth terms in equation (\ref{spd}) is to introduce an operator identity in order to get rid off the multiplication of the resolvents in the residue integrals such that all residue integrals contain just operators commuting with each other with arbitrary powers of them. It turns out that the same result which has been found with the assistance of the Feynman parametrization will be obtained as it should be. This is possible with the following operator identity
\begin{eqnarray}
\frac{1}{(\lambda + \tilde{A})^3} \frac{1}{(\lambda + \tilde{A}^\ast)} &=& \frac{1}{8} \frac{1}{( \lambda + \tilde{\phi} )^3} \frac{1}{(\lambda + \tilde{A})} + \frac{1}{8} \frac{1}{( \lambda + \tilde{\phi} )^3} \frac{1}{(\lambda + \tilde{A}^\ast)} \nonumber \\
& & + \frac{1}{4} \frac{1}{( \lambda + \tilde{\phi} )^2} \frac{1}{(\lambda + \tilde{A})^2}+ \frac{1}{2} \frac{1}{(\lambda + \tilde{\phi})} \frac{1}{(\lambda + \tilde{A})^3} \,.
\end{eqnarray}
However, if one pays more attention to the expression above, it is easy to notice that all the terms have a factor which can be located on the cut. On account of that, a complex parameter is supposed to be introduced to the symbol of the scalar field $\tilde{\phi}$, so it becomes $\tilde{\phi} + i \varepsilon$, where $\varepsilon$ is positive but small and will eventually be made to approach zero, that is,
\[
\tilde{C} = \tilde{\phi} + i p \cdot \gamma + i \varepsilon \quad , \quad \tilde{C}^\prime = \tilde{\phi} - i p \cdot \gamma + i \varepsilon \quad,\quad \tilde{B} = \tilde{\phi} + i \varepsilon
\]
\begin{equation}
\tilde{A} = \lim_{\varepsilon \rightarrow 0^+} \tilde{C} \quad,\quad \tilde{A}^\ast = \lim_{\varepsilon \rightarrow 0^+} \tilde{C}^\prime \quad,\quad \tilde{\phi} = \lim_{\varepsilon \rightarrow 0^+} \tilde{B} \,.
\end{equation}
By means of these new operators with the operator identity already given, the following integral can be given as,
\begin{eqnarray}
\frac{\sin \pi s}{\pi} \int^\infty_0 d \lambda \lambda^{-s} \frac{1}{(\lambda + \tilde{A})^3} \frac{1}{(\lambda + \tilde{A}^\ast)} = \lim_{\varepsilon \rightarrow 0^+} \frac{\sin \pi s}{\pi} \int^\infty_0 d \lambda \lambda^{-s} \frac{1}{(\lambda + \tilde{C})^3} \frac{1}{(\lambda + \tilde{C}^\prime)} \nonumber \\
= \lim_{\varepsilon \rightarrow 0^+} \frac{\sin \pi s}{\pi} \int^\infty_0 d \lambda \lambda^{-s} \frac{1}{8} \frac{1}{( \lambda + \tilde{B} )^3} \frac{1}{(\lambda + \tilde{C})} + \lim_{\varepsilon \rightarrow 0^+} \frac{\sin \pi s}{\pi} \int^\infty_0 d \lambda \lambda^{-s} \frac{1}{8} \frac{1}{( \lambda + \tilde{B} )^3} \frac{1}{(\lambda + \tilde{C}^\prime)} \nonumber \\ + \lim_{\varepsilon \rightarrow 0^+} \frac{\sin \pi s}{\pi} \int^\infty_0 d \lambda \lambda^{-s} \frac{1}{4} \frac{1}{( \lambda + \tilde{B} )^2} \frac{1}{(\lambda + \tilde{C})^2} + \lim_{\varepsilon \rightarrow 0^+} \frac{\sin \pi s}{\pi} \int^\infty_0 d \lambda \lambda^{-s}  \frac{1}{2} \frac{1}{( \lambda + \tilde{B} )} \frac{1}{(\lambda + \tilde{C})^3} \,.
\end{eqnarray}
After taking these residue integrals with the $\varepsilon$ small limit afterwards, the above integral is given by
\begin{eqnarray}
\frac{\sin \pi s}{\pi} \int^\infty_0 d \lambda \lambda^{-s} \frac{1}{(\lambda + \tilde{A})^3} \frac{1}{(\lambda + \tilde{A}^\ast)} &=& - \frac{1}{8} \left[ \frac{1}{(i p \cdot \gamma)^3} \tilde{A}^{-s} - \frac{1}{(i p \cdot \gamma)^3} \tilde{A^\ast}^{-s} \right] \nonumber \\ - \frac{s}{4 (i p \cdot \gamma)^2} \tilde{A}^{-s-1} &-& \frac{s (s+1)}{4 (i p \cdot \gamma)} \tilde{A}^{-s-2} \,.
\end{eqnarray}
The second term is, then,
\begin{eqnarray}\label{2ndterm}
- \frac{\partial_{\mu \nu} \tilde{\phi}}{4} \mathrm{tr} \int \frac{d^4 p}{(2 \pi)^4} \frac{\sin \pi s}{\pi} \int^\infty_0 d \lambda \lambda^{-s} \frac{1}{(\lambda + \tilde{A})^3} \frac{1}{(\lambda + \tilde{A}^\ast)} \gamma^\mu \gamma^\nu = \nonumber \\ + \frac{\partial^2 \phi}{32} \mathrm{tr} \int \frac{d^4 p}{(2 \pi)^4} \frac{i p \cdot \gamma}{p^4} \left( \tilde{A}^{-s} - \tilde{A^\ast}^{-s}\right) - \frac{s \partial^2 \phi}{16} \mathrm{tr} \int \frac{d^4 p}{(2 \pi)^4} \frac{1}{p^2} \tilde{A}^{-s-1} \nonumber \\ - \frac{s (s+1) \partial^2 \phi}{16} \mathrm{tr} \int \frac{d^4 p}{(2 \pi)^4} \frac{i p \cdot \gamma}{p^2} \tilde{A}^{-s-2} \,.
\end{eqnarray}
To be explicit, the first integral can be calculated in $d$--dimensions and then one lets $d \rightarrow 4$. This is done as follows,
\begin{eqnarray}
\mathrm{tr} \int\frac{d^d p}{(2 \pi)^d} \frac{i p \cdot \gamma}{p^4} \left[ \frac{1}{(\tilde{\phi} + i p \cdot \gamma)^s} - \frac{1}{(\tilde{\phi} - i p \cdot \gamma)^s} \right] \nonumber \\ =  \frac{4 s}{\Gamma(d/2)} \frac{\tilde{\phi}^{-s-1}}{( 2 \pi)^{d/2}} \int^\infty_0 d p \, p^{d-3} {}_2 F_1 \left( \frac{s+1}{2} , \frac{s+2}{2} ; \frac{3}{2} ; - \frac{p^2}{\tilde{\phi}^2} \right) \nonumber \\  = \frac{2 s}{\Gamma(d/2)} \frac{\tilde{\phi}^{-s-1}}{( 2 \pi)^{d/2}} \frac{\Gamma(3/2) \Gamma(d/2-1)}{\Gamma(5/2-d/2)} \frac{\Gamma(s+3-d)}{\Gamma(s+1)} 2^{d-2} \left( \tilde{\phi}^2 \right)^{d/2-1} \,,
\end{eqnarray}
if $ 0 < \Re(d/2-1) < \Re(s/2 + 1/2 ) , \Re (s/2+1) $ and $ \left| \arg(1/ \tilde{\phi}^2) \right|< \pi $.
The first term in equation (\ref{2ndterm}) becomes the following after letting $d \rightarrow 4$ in the above $d$ dimensional expression,
\begin{equation}
\frac{\partial^2 \phi}{32} \mathrm{tr} \int \frac{d^4 p}{(2 \pi)^4} \frac{i p \cdot \gamma}{p^4} \left( \tilde{A}^{-s} - \tilde{A^\ast}^{-s}\right) = \frac{\partial^2 \tilde{\phi}}{32 \pi^2} \frac{\phi^{1-s}}{(s-1)} \,.
\end{equation}
The other terms in equation (\ref{2ndterm}) can be calculated in the same manner and one gets
\begin{eqnarray}
- \frac{s \partial^2 \phi}{16} \mathrm{tr} \int \frac{d^4 p}{(2 \pi)^4} \frac{1}{p^2} \tilde{A}^{-s-1} &=& \frac{\partial^2 \tilde{\phi}}{32 \pi^2} \frac{\phi^{1-s}}{(s-1)} \\
-\frac{s (s+1) \partial^2 \phi}{16} \mathrm{tr} \int \frac{d^4 p}{(2 \pi)^4} \frac{i p \cdot \gamma}{p^2} \tilde{A}^{-s-2} &=& \frac{\partial^2 \tilde{\phi}}{16 \pi^2} \frac{\phi^{1-s}}{(s-1)} \,.
\end{eqnarray}
The second term in the expansion is, then,
\begin{equation}
- \frac{\partial_{\mu \nu} \tilde{\phi}}{4} \mathrm{tr} \int \frac{d^4 p}{(2 \pi)^4} \frac{\sin \pi s}{\pi} \int^\infty_0 d \lambda \lambda^{-s} \frac{1}{(\lambda + \tilde{A})^3} \frac{1}{(\lambda + \tilde{A}^\ast)} \gamma^\mu \gamma^\nu = \frac{\partial^2 \tilde{\phi}}{8 \pi^2} \frac{\phi^{1-s}}{(s-1)} \,,
\end{equation}
which is the same result already found with the assistance of the Feynman parametrization.

The next thing should be done is to do the same tedious calculations for the fourth term in the expansion (\ref{spd}). Using the operator identity in the semigroup integral representation, this term can be given by
\begin{eqnarray}
\frac{\partial_{\mu \nu} \tilde{\phi}}{4} \mathrm{tr} \int \frac{d^4 p}{(2 \pi)^4} \frac{\sin \pi s}{\pi} \int^\infty_0 d \lambda \lambda^{-s} \frac{1}{(\lambda + \tilde{A})^3} \frac{1}{(\lambda + \tilde{A}^\ast)} \frac{p^\mu}{p^2} p \cdot \gamma \gamma^\nu \nonumber \\
= \frac{\partial_{\mu \nu} \tilde{\phi}}{32} \mathrm{tr} \int \frac{d^4 p}{(2 \pi)^4} \frac{p^\mu}{p^6} i p \cdot \gamma \left( \tilde{A^\ast}^{-s} - \tilde{A}^{-s}\right) p \cdot \gamma \gamma^\nu \nonumber \\ + \frac{s \partial_{\mu \nu} \tilde{\phi}}{16} \mathrm{tr} \int \frac{d^4 p}{(2 \pi)^4} \frac{p^\mu}{p^4} \tilde{A}^{-s-1} p \cdot \gamma \gamma^\nu + \frac{s (s+1) \partial_{\mu \nu} \tilde{\phi}}{16} \mathrm{tr} \int \frac{d^4 p}{(2 \pi)^4} \frac{p^\mu}{p^4} i p \cdot \gamma \tilde{A}^{-s-2} p \cdot \gamma \gamma^\nu \,.
\end{eqnarray}
We will just give the results for the momentum integrals since the calculations are the same as the ones which have been done for the second term in the expansion (\ref{spd}),
\begin{eqnarray}
\frac{\partial_{\mu \nu} \tilde{\phi}}{32} \mathrm{tr} \int \frac{d^4 p}{(2 \pi)^4} \frac{p^\mu}{p^6} i p \cdot \gamma \left( \tilde{A^\ast}^{-s} - \tilde{A}^{-s}\right) p \cdot \gamma \gamma^\nu &=& - \frac{\partial^2 \tilde{\phi}}{128 \pi^2} \frac{\tilde{\phi}^{1-s}}{s-1} \,,\\
\frac{s \partial_{\mu \nu} \tilde{\phi}}{16} \mathrm{tr} \int \frac{d^4 p}{(2 \pi)^4} \frac{p^\mu}{p^4} \tilde{A}^{-s-1} p \cdot \gamma \gamma^\nu &=& - \frac{\partial^2 \tilde{\phi}}{128 \pi^2} \frac{\tilde{\phi}^{1-s}}{s-1} \,,\\
\frac{s (s+1) \partial_{\mu \nu} \tilde{\phi}}{16} \mathrm{tr} \int \frac{d^4 p}{(2 \pi)^4} \frac{p^\mu}{p^4} i p \cdot \gamma \tilde{A}^{-s-2} p \cdot \gamma \gamma^\nu &=& - \frac{\partial^2 \tilde{\phi}}{64 \pi^2} \frac{\tilde{\phi}^{1-s}}{s-1} \,.
\end{eqnarray}
The fourth term is, therefore,
\begin{equation}
\frac{\partial_{\mu \nu} \tilde{\phi}}{4} \mathrm{tr} \int \frac{d^4 p}{(2 \pi)^4} \frac{\sin \pi s}{\pi} \int^\infty_0 d \lambda \lambda^{-s} \frac{1}{(\lambda + \tilde{A})^3} \frac{1}{(\lambda + \tilde{A}^\ast)} \frac{p^\mu}{p^2} p \cdot \gamma \gamma^\nu = - \frac{\partial^2 \tilde{\phi}}{32 \pi^2} \frac{\tilde{\phi}^{1-s}}{s-1} \,.
\end{equation}
If these terms are collected together, the $\hbar^2$--order correction in the semiclassical expansion is found to be equal to,
\begin{eqnarray}
- \frac{\partial_{\mu \nu} \tilde{\phi}}{4} \mathrm{tr} \int \frac{d^4 p}{(2 \pi)^4} \frac{\sin \pi s}{\pi} \int^\infty_0 d \lambda \lambda^{-s} \frac{1}{(\lambda + \tilde{A})^3} \frac{1}{(\lambda + \tilde{A}^\ast)} \gamma^\mu \gamma^\nu \nonumber \\ - \frac{\partial_{\mu \nu} \tilde{\phi}}{4} \mathrm{tr} \int \frac{d^4 p}{(2 \pi)^4} \frac{\sin \pi s}{\pi} \int^\infty_0 d \lambda \lambda^{-s} \frac{1}{(\lambda + \tilde{A})^4} \frac{p^\mu}{p^2} p \cdot \gamma \gamma^\nu \nonumber \\
+ \frac{\partial_{\mu \nu} \tilde{\phi}}{4} \mathrm{tr} \int \frac{d^4 p}{(2 \pi)^4} \frac{\sin \pi s}{\pi} \int^\infty_0 d \lambda \lambda^{-s} \frac{1}{(\lambda + \tilde{A})^3} \frac{1}{(\lambda + \tilde{A}^\ast)} \frac{p^\mu}{p^2} p \cdot \gamma \gamma^\nu
&=& \frac{\partial^2 \tilde{\phi}}{16 \pi^2} \frac{\tilde{\phi}^{1-s}}{s-1} \,, \nonumber \\
\end{eqnarray}
which is exactly the same result that we have found by means of the Feynman parametrization.

Thus the semiclassical expansion of the zeta function and the determinant of the Dirac operator with a scalar field $\phi$ can, respectively, be given by
\begin{eqnarray}
\zeta\left( s | \gamma \cdot \partial + \phi \right) &=& \frac{1}{\pi^2} \int d^4x \left[  3 \tilde{\phi}^{4-s} \frac{\Gamma(s-4)}{\Gamma(s)} + \frac{1}{16} \tilde{\phi}^{1-s} \frac{\Gamma(s-1)}{\Gamma(s)} \partial^2 \tilde{\phi} \right] + \cdots \,, \nonumber \\ \\
\ln \det (\gamma \cdot \partial + \phi) &=& \frac{1}{16 \pi^2} \int d^4x \left[ \phi^4 \ln \left( \frac{\phi^2}{\mu^2} e^{-25/6}\right) + \ln \left( \frac{\phi^2}{\mu^2} \right) \frac{1}{2} \partial_\mu \phi \partial^\mu \phi \right] + \cdots \,, \nonumber \\
\end{eqnarray}
which agrees with the result given in \cite{perry} up to a numerical factor.
\section{The large--$N$ Yukawa theory}\label{5thsec}
In this section a set of $N+1$ massless Dirac fermions that are both $U(N+1)$ symmetric and couples to a scalar field via the well known Yukawa coupling is studied at large--$N$ regime. The self coupling of the scalar field is left unspecified. The Euclidean action of this theory, similiar to the ones in refrences \cite{zinn1,zinn2}, is given by
\begin{equation}
S \left[ \bar{\Psi} , \Psi , \phi \right] = \int d^4 x \left\{ - \bar{\Psi} \left( \gamma \cdot \partial + g \phi \right) \Psi + \frac{1}{2} \partial_\mu \phi \partial^\mu \phi + N V \left[ \frac{\phi^2}{N} \right] \right\} \,.
\end{equation}
For the large--$N$ analysis, $N$ fermion fields are integrated out and the fields $\psi_{N+1}$, $\bar{\psi}_{N+1}$ and $\phi$ are rescaled as $\sqrt{N} \psi_{N+1}$, $\sqrt{N} \bar{\psi}_{N+1}$ and $\phi / g$, respectively, with a redefined coupling constant $\tilde{g}^2 = g^2 N$ which is kept fixed as $N \rightarrow \infty$. In this way, one gets
\begin{eqnarray}\label{larnyuk}
S_{eff} \left[ \bar{\psi} , \psi , \phi \right] &=& N \int d^4 x \left\{ - \bar{\psi} \left( \gamma \cdot \partial + \phi \right) \psi + \frac{1}{2 \tilde{g}^2} \partial_\mu \phi \partial^\mu \phi + V \left[ \frac{\phi^2}{\tilde{g}^2} \right] \right\} \nonumber \\
&& - N \ln \det (\gamma \cdot \partial + \phi)\,.
\end{eqnarray}
In the limit $N \rightarrow \infty$, the contributions coming from the extremals of this action dominates the functional integral. The functional to zeroth order in $1/N$ is just the $N=\infty$ quantum effective action without any corrections coming from the next $1/N$ orders. If one plugs the semiclassical expansion for the determinant of the same Dirac operator, which has been calculated in the previous section, into equation (\ref{larnyuk}), the $N=\infty$ quantum effective action is, thus, given by
\begin{eqnarray}\label{effact}
\Gamma_0 \left[ \bar{\psi} , \psi , \phi \right] &=& \int d^4 x  \left\{ - \bar{\psi} \left( \gamma \cdot \partial + \phi \right) \psi + \left[ \frac{1}{\tilde{g}^2}   - \frac{1}{16 \pi^2} \ln \left( \frac{\phi^2}{\mu^2} \right) \right]  \frac{1}{2} \partial_\mu \phi \partial^\mu \phi \right. \nonumber \\
&& \left. + V \left[ \frac{\phi^2}{\tilde{g}^2} \right] - \frac{\phi^4}{16 \pi^2} \ln \left( \frac{\phi^2}{\mu^2} e^{-25/6}\right) \right\} \,.
\end{eqnarray}
If one pays more attention to this effective action, it is easy to notice that there is a term involving gradients which is multiplied by a logarithm of the field. This combination, of course, arises from the fact that the scalar field is assumed to be inhomogeneous or in other words nonconstant while the determinant coming from integrating the $N$ fermions out is being calculated. Here it is appropriate to make a comparison between the $\phi^4$--theory and the Yukawa theory as far as the perturbation theory in terms of $\varepsilon$--expansion is concerned. What is known from the perturbation theory is that there is no one--loop wave function renormalization for the $\phi^4$--theory whereas one has to deal with that for the Yukawa theory even at one loop. The renormalization constants for the fields as well as for the other parameters in the theory should be defined in terms of the poles of $\varepsilon$ and appropriate counterterms should also be introduced in order to end up with a finite set of parameters in the theory. Although one does not have to worry about wave function renormalization to first order in $\phi^4$--theory, the kinetic term is multiplied by a nontrivial functional of the field \cite{itzykson}. It is easy to obtain this factor by replacing $V(x)$ with $m^2 + \lambda \phi^2(x)/2$ in equation (\ref{bosdet}) and this yields
\begin{equation}
\frac{\lambda^2}{6 (4 \pi)^2}\frac{\phi^2}{\left(2 m^2 + \lambda \phi^2\right)} \,,
\end{equation}
which is actually the first nontrivial term of the $Z_{eff}(\phi)$, which is the functional multiplier of the kinetic term in the
expansion of the effective action in terms of the field itself.
However, to second order in the loop expansion there is a wave
function renormalization in the $\phi^4$--theory and the terms
involving gradients multiplied by logarithms of the field actually
occur as in the Yukawa theory. Therefore it can presumably be
thought that the occurrence of that type of kinetic terms with
logarithmic multipliers in the effective action expansion is a
manifestation of the divergences encountered in parameter
renormalization for the fields themselves. Thus the reason that
the kinetic term multiplied by a logarithm of the field is
encountered in the Yukawa theory even in the first order in the
loop expansion as in equation (\ref{effact}) arises from the fact
that there is an inevitable wave function renormalization in the
Yukawa theory at one loop. As it is written, this theory looks
unstable for the different regimes for the scalar fields, however
one has to find the physical fields with the correct classical
configuration and then define the renormalized action afterwards.
\section{Conclusion}
In this paper, what is studied basically depends on developing an
alternative and powerful method to give a semiclassical expansion
for infinite dimensional determinants encountered frequently in
quantum field theory applications. In order to establish this,
Weyl symbol calculus, semigroup integral representation of
operators and zeta function regularization techniques are
extensively used. For Weyl type symbol calculus there is a one to
one correspondence between the operators acting on the Hilbert
space, whose determinants are needed for the quantum effective
action calculations in terms of loops. Although the main idea of
this method, which is finding an expansion for the arbitrary complex
powers of elliptic differential operators in pseudo differential
operator language, goes back to Seeley's excellent paper
\cite{seeley}, a slight modification is needed so as to obtain the
logarithmic terms which are highly important for the quantum field
theoretical calculations through the expansion for the zeta
function regularized determinants. In order to calculate the
semiclassical expansion for determinants, allowing us to obtain
logarithmic terms for the operators under consideration, not only
the momentum part which is formerly introduced as the principal
symbol in these kind of calculations, but also the so called
potential part of the symbol of the operator should be kept, that
is, both the momentum and the potential term should be recognized
as the principal symbol of the pseudo differential operator. Power
of this method also comes from the fact that arbitrary
differential operators which contain different types of matrix
operators can easily be treated in this framework, since the
operators in the Hilbert space transform into matrix valued
symbols in such cases and the same recursive expansions arising
from the product rule between the phase space functions are still
valid as long as the matrix degrees of freedoms of the symbols are
properly taken into account.

In our opinion not only this method for calculating the
regularized determinants for elliptic operators is more natural
but also more systematic regarding the corrections involving
higher order derivatives. The application of the renormalization
group equations to the large--$N$ Yukawa theory within this formalism is left to the
future works.

\section{Acknowledgement}
Burak Tevfik Kaynak would like to thank Cemsinan Deliduman and Tongu\c{c} Rador for useful discussions.


\begin{thebibliography}{99}
\bibitem{ramond} P. Ramond, \emph{Field Theory: A Modern Primier}, (Addison--Wesley, 1990).
\bibitem{sweinberg} S. Weinberg, \emph{The Quantum Theory of Fields Vol. II}, (Cambridge University Press, 2005).
\bibitem{seeley} R. T. Seeley, \emph{Complex Powers of an Elliptic Operator}, Amer. Math. Soc. Proc. Symp. Pure Math. {\bf 10}, 288 (1967)
\bibitem{singer1} D. B. Ray and I. M. Singer, \emph{R--torsion and the Laplacian on Riemannian manifolds}, Adv. in Math. {\bf 7}, 145 (1971).
\bibitem{singer2} D. B. Ray and I. M. Singer, \emph{Analytic torsion for complex manifolds}, Annals of Math. {\bf 98}, 154 (1973).
\bibitem{hawking} S. Hawking, \emph{Zeta Function Regularization of Path Integrals in Curved Space--Time}, Commun. Math. Phys. {\bf 55}, 133 (1977).
\bibitem{ball} R. D. Ball, \emph{Chiral Gauge Theory}, Phys. Rep. {\bf 182}, 1 (1989).
\bibitem{camporesi} R. Camporesi, \emph{Harmonic analysis and propagators on homogeneous spaces}, Phys. Rept. {\bf 196}, 1 (1990).
\bibitem{bytsenko} A. A. Bytsenko, G. Congola, L. Vanzo and S. Zerbini, \emph{Quantum fields and extended objects in spacetimes with constant curvature spatial section}, Phys. Rept. {\bf 266}, 1 (1996).
\bibitem{zinn1} J. Zinn--Justin, \emph{Quantum Field Theory and Critical Phenomena}, (Clarendon Press, 1996).
\bibitem{kontsevich} M. Kontsevich and S. Vishik, \emph{Determinants of elliptic pseudodifferential operators}, MPI-94-30, hep-th/9404046.
\bibitem{elizalde} G. Cognola, E. Elizalde and S. Zerbini, \emph{Dirac Functional Determinants in Terms of the Eta Invariant and the Noncummutative Residue}, Commun. Math. Phys. {\bf 237}, 507 (2003).
\bibitem{zerbini} S. Zerbini, \emph{The Multiplicative Anomaly of Regularized Funtional Determinants}, Nucl. Phys. Proc. Suppl. {\bf 104}, 224 (2002).
\bibitem{shubin} M. A. Shubin and S. I. Anderson, \emph{Pseudodifferential Operators and Spectral Theory}, (Springer--Verlag, 1987).
\bibitem{saravi1} H. Falomir, R. E. Gamboa Saravi, M. A. Muschietti, E. M. Santangelo and J. E. Solomin, \emph{Determinants of Dirac operators with local boundary conditions}, hep-th/9608101.
\bibitem{gelfand} I. M. Gelfand and A. M. Yaglom, \emph{Integration In Functional Spaces And Its Applications In Quantum Physics}, J. Math. Phys. {\bf 1}, 48 (1960).
\bibitem{kirsten1} K. Kirsten and A. J. McKane, \emph{Functional determinants by contour integration methods}, Annals of Phys. {\bf 308}, 502 (2003).
\bibitem{kirsten2} K. Kirsten and A. J. McKane, \emph{Funtional determinants for general Sturm--Liouville problems}, J. Phys. {\bf A37}, 4649 (2004).
\bibitem{dunne} G. V. Dunne and K. Kirsten, \emph{Funtional determinants for radial operators}, J. Phys. {\bf A39}, 11915 (2006).
\bibitem{hormander} L. Hormander, \emph{The Weyl Calculus of Pseudo--Differential Operators}, Comm. Pure Appl. Math. {bf 32}, 359 (1979)
\bibitem{folland} G. B. Folland, {\it Harmonic Analysis in Phase Space},
(Princeton, 1989).
\bibitem{kirsten3} M. Bordag, B. Geyer, K. Kirsten and E. Elizalde, \emph{Zeta function determinant of the Laplace operator on the $D$--dimensional ball}, Commun. Math. Phys. {\bf 179}, 215 (1996).
\bibitem{kirsten4} M. Bordag, K. Kirsten and J. S. Dowker, \emph{Heat--kernels and functional determinants on the generalized cone}, Commun. Math. Phys. {\bf 182}, 371 (1996).
\bibitem{kirsten5} K. Kirsten, P. Loya and J. Park, \emph{Zeta functions of Dirac and Laplace-type operators over finite cylinders}, Annals of Phys. {\bf 321}, 1814 (2006).
\bibitem{saravi2} R. E. Gamboa Saravi, M. A. Muschietti, F. A. Schaposnik and J. E. Solomin, \emph{Chiral Symmetry and Functional Integral}, Annals of Phys. {\bf 157}, 360 (1984).
\bibitem{salcedo} L. L. Salcedo and E. R. Arriola, \emph{Wigner Transformation for The Determinant of Dirac Operators}, Annals of Phys. {\bf 250}, 1 (1996).
\bibitem{langmann} E. Langmann, \emph{Generalized Yang--Mills Actions From Dirac Operator Determinants}, J. MAth. Phys. {\bf 42}, 5238 (2001).
\bibitem{schwarz} A. S. Schwarz, \emph{Quantum Field Theory and Topology}, (Springer, 1993).
\bibitem{rajeev} S. G. Rajeev, \emph{New Classical Limits of Quantum Theories}, hep-th/0210179.
\bibitem{wojciechowski} J. Park and K. P. Wojciechowski, \emph{Analytic Surgergy of the $\zeta$--determinant of the Dirac Operator}, Nucl. Phys. Proc. Supp. {\bf B104}, 89 (2002).
\bibitem{itzykson} J. Illiopoulos, C. Itzykson and A. Martin, \emph{Functional Methods and Perturbation Theory}, Rev. Mod. Phys. {\bf 47}, 165 (1975).
\bibitem{chan} L. Chan, \emph{Effective--Action Expansion in Perturbation Theory}, Phys. Rev Let. {\bf 54}, 1222 (1985).
\bibitem{eweinberg} S. Coleman and E. Weinberg, \emph{Radiative Corrections as the Origin of Spontaneous Symmetry Breaking}, Phys. Rev. {\bf D7}, 1888 (1973).
\bibitem{prudnikov} A.P. Prudnikov, Yu. A. Brychkov and O.I. Marichev, \emph{Integrals and series}, (CRC, 1990).
\bibitem{perry} M. Li and R. J. Perry, \emph{Calculating Boson and Fermion Loops in $3+1$ Dimensions and The Derivative Expansion}, Phys. Rev. {\bf D37}, 1670 (1988).
\bibitem{zinn2} J. Zinn--Justin, \emph{Four--Fermi Interaction Near Four Dimensions}, Nucl. Phys. {\bf B367}, 105 (1991).
\end{thebibliography}
\end{document}